\documentstyle[emulateapj,onecolfloat]{article}

\input epsf
 at 14.40truept
 at 17.28truept
\newcount\itemno
\def \ino         { \the\itemno\global\advance\itemno by 1 }

\def \ie	{\hbox{\it i.e.}}

\def \kev	{{\rm\ keV}}
\def \kb	{{\rm\ k}}
\def \temp	{{\rm T}}
\def \msol	{{\rm M}_\odot}

\def \h50inv	{\hbox{$\, h^{-1}_{50}$} }

\def \ssim	{\! \sim \!}

\def\apj{{\it Ap. J.\/}}

\def\mnras{{\it M.N.R.A.S.\/}}

\def\\{\hfil\break}
\def\spose#1{\hbox to 0pt{#1\hss}}
\def\lta{\mathrel{\spose{\lower 3pt\hbox{$\mathchar"218$}}
     \raise 2.0pt\hbox{$\mathchar"13C$}}}
\def\gta{\mathrel{\spose{\lower 3pt\hbox{$\mathchar"218$}}
     \raise 2.0pt\hbox{$\mathchar"13E$}}}

\def\beq{\begin{equation}}
\def\eeq{\end{equation}}
\font\tfa=cmr10 at 8.00pt
\tfa
\lefthead{Metzler}
\righthead{The Cluster SZ --- Mass Correlation}

\begin{document}
\twocolumn[
\title { The Cluster SZ --- Mass Correlation }

\author {Christopher A. Metzler}
\affil {Loomis Laboratory of Physics, University of Illinois,
1110 W. Green Street, Champaign, IL 61801--3080 USA}
\affil{and}
\affil {NASA/Fermilab Astrophysics Group, Fermi National Accelerator
Laboratory, Box 500, Batavia, IL 60510-0500 USA}

\begin{abstract}
N-body + hydrodynamic simulations of galaxy clusters are used to
demonstrate a correlation between galaxy cluster
mass and the strength of the Sunyaev--Zel'dovich (SZ) effect induced by
the cluster.  The intrinsic scatter in the correlaton is larger
than seen in the cluster mass --- X--ray temperature correlation,
but smaller than seen in the correlation between mass (or temperature)
and X--ray luminosity, as expected.  Using the
convergence to self--similarity of cluster structure
at larger radii, a simple area--averaged SZ value derived from
mock SZ maps also correlates well with mass; the intrinsic scatter
in this correlation is comparable to that seen in simulations
for the mass --- temperature correlation.  Such a relation may
prove a powerful tool for estimating cluster masses at higher
redshifts.
\end{abstract}

]

\section{Motivation}

Correlations between galaxy cluster mass and
observables such as X--ray luminosity $L_x$  or temperature $\temp$
are used extensively to probe cosmology.  For instance, the abundance
of clusters as a function of temperature has been used
to constrain the normalization of the power spectrum of primordial
fluctuations, $\sigma_8$, while searches for evolution in the
cluster X--ray luminosity function have been employed to argue
for values of the cosmological density parameter $\Omega_0$.
Theoretical models such as the Press--Schechter (\cite{PS74} 1974)
formalism characteristically
provide the abundance of clusters as a function of the cluster
mass.  The use of temperature or luminosity functions to constrain
the parameters in the theoretical model therefore depends upon
a relation between $\temp$ or $L_x$ and mass.
Theoretical arguments and cluster simulations indicate that
temperature and X--ray luminosity should have a power law
dependence on mass,
while observations of cluster abundances and correlations between
observables
are consistent with such simple power law relations holding true
for galaxy clusters.

A comparatively new tool for investigating clusters is provided by
observations of the Sunyaev--Zel'dovich (SZ) effect
(\cite{SZ72} 1972; \cite{Birk98} 1998) ---
the distortion in the cosmic microwave background
spectrum produced by Compton upscattering of microwave background photons
by electrons in the hot intracluster medium (ICM).
The SZ effect has two properties that make it particularly interesting
for examining high--redshift systems.  First, the signal strength
is not attenuated by redshift, in contrast with X--ray flux.  Second,
for clusters of
fixed mass, self--similar scaling relations predict
a strong positive evolution with redshift.  Since observations of
X--ray temperature, X--ray surface brightness, and the SZ effect
depend upon the density and temperature structure of cluster gas in different
ways, the SZ effect gives a complimentary probe of the state of
the ICM.

An interesting question is whether the strength of the SZ effect
correlates with cluster mass, as is expected of $\temp$
or $L_x$.  A simple argument reproduced below suggests
that such a correlation
should also be present; verification of this would provide
a consistency check upon the simple theoretical models used to
describe clusters.  Furthermore,
accurate X--ray temperature determinations become harder with
increasing redshift.  As the strength of the
SZ signal is redshift--independent, and as clusters of a fixed
mass produce a stronger effect at higher redshift, a correlation
between SZ signal and mass yields a powerful method for estimating
cluster masses at redshifts beyond those currently probed by X--ray
temperature estimates.

In this {\em Letter}, we use simulated clusters to examine the relation
between cluster mass and the SZ effect.  We construct mock SZ
images of the simulated clusters, extract SZ observables, and examine
correlations with mass as well as scatter about the correlations.
We find that the central SZ signal exhibits the expected self--similar
scaling with mass in these simulations, with a scatter significantly
smaller than is seen in simulated $L_x$--$M$ relations or the observed
cluster $L_x$--$\temp$ relation.  We will also note that a simple
area--averaging of the SZ signal reduces the dispersion to a level
near that expected from the cluster
X--ray temperature.  Some concerns about applying this result, as well
as directions for future study, will be described at the end.

% \vfill\eject

\section{Models}

The simulated clusters were produced using the N--body + hydrodynamic
code P3MSPH; both the simulation code and analysis techniques are
described in \cite{Evra90b} (1990).  Each simulation run produced an individual
cluster, with initial conditions drawn from the $\Omega\,=\,1$,
$\Omega_b\,=\,0.1$, $h\,=\,0.5$ (assumed throughout this paper),
$\sigma_8\,=\,0.6$ cold dark matter model using the constrained--realization
technique of Bertschinger (1987).  The clusters simulated ranged in mass
from $8\times 10^{13}$ to $1.1\times 10^{15}\msol$, with ICM
temperatures between $0.7$ and $4.3\kev$.
A total of 65536 particles were used in each simulation,
half in dark matter and half in baryons.  The fractional mass resolution
was comparable amongst the simulated clusters, with around 3500 gas
particles typically residing in the cluster at $z\,=\,0$.  A total of
73 simulated clusters were produced in this fashion.
The final, $z\,=\,0$ configuration of each simulated cluster was
then ``imaged'' in the SZ by producing a pixel map of values of the
Compton y--parameter, each pixel's value constructed by a line--of--sight
pressure integral through the simulated cluster.  For each run, an
SZ map was produced in three orthogonal directions; this produced
a total of 219 maps.  In extracting
a central value for the y--parameter, $y_{0}$, the cluster ``center''
is defined by the projected location of the bottom of the cluster
potential well, which traces well the projected location
of the maximum value of the y--parameter.

\section{The Cluster y--M Relation}

The Compton y--parameter induced by upscattering
of microwave background photons along a line of sight $\ell$ through
the cluster is given by
\beq
y \,=\, \int n_{e}\left(\frac{\kb\temp}{m_{e}c^{2}}\right)\sigma_{T} \rm{d}\ell.
\eeq
This equation
can be used to derive a self--similar scaling relation for clusters
which are near isothermality when density--weighted.  Writing the
density in terms of the background density $\bar{\rho}$,
expressing length scales in terms
of the radius $r_{\delta}$ at some fixed overdensity $\delta$,
and factoring out the temperature, we have
\beq
y \,\propto\, \bar{\rho}\ r_{\delta} \temp
\int\left(\frac{\rho_{gas}}{\bar{\rho}}\right)
\rm{d}\left(\frac{\ell}{r_{\delta}}\right).
\eeq
If clusters are self--similar, then the integral produces a number which
is identical for clusters of any mass (\cite{NFW95} 1995; \cite{Metz95} 1995;
\cite{MeEv98} 1998), and varies only with the
choice of line of sight.  With
$M\,=\,\delta\bar{\rho}\left(4\pi/3\right)r_{\delta}^{3}$,
$\bar{\rho}\,\propto\,\left(1\,+\,z\right)^{3}$, and the virial
scaling $\temp\,\propto\,M^{2/3}\left(1\,+\,z\right)$, we have
\beq
\label{eq:ymreln}
y\,\propto\,M \left(1\,+\,z\right)^{3}.
\eeq
This yields the scaling, with mass and redshift, of the y--parameter
induced along a line of sight with a 2--D radius equal to the 3--D
radius of some fixed fiducial overdensity; the constant of proportionality
depends upon the overdensity in question.  In particular,
the values of the SZ central temperature decrement should scale in
this fashion.

Do the simulated clusters obey this relation?  We have 219 pairs
of $y_{0}$ and mass within an overdensity of 200
with which to test the relation; the results are plotted in
Figure 1.  The best--fit power law relation is plotted as a solid
line; the best--fit power law slope is
% $1.01\pm 0.05$,
$0.98\pm 0.04$,
well in
agreement with the theoretical expectation, Eq. (\ref{eq:ymreln}).
Demanding Eq. (\ref{eq:ymreln}) be exactly correct yields a
best--fit
coefficient of proportionality which is not significantly different
from that found in the general power--law fit.

\begin{figure}[t]
\centerline{\epsfxsize=3.5truein\epsffile{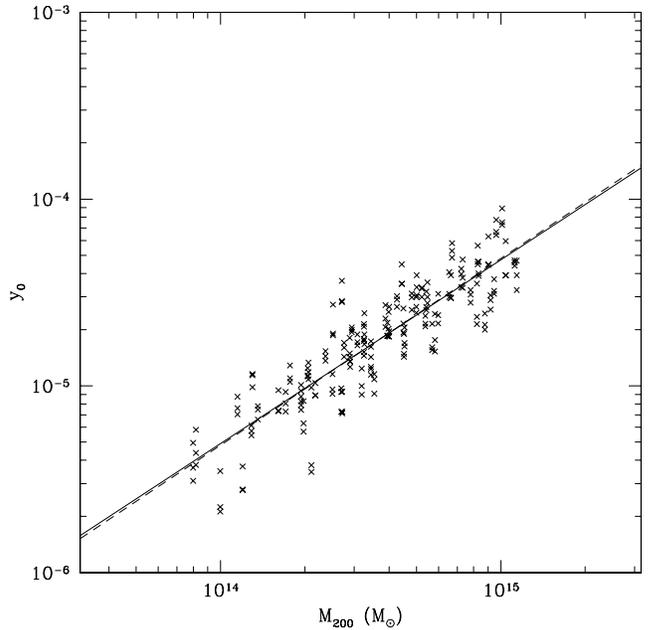}}
% \centerline{\epsffile{fig1.ps}}
\caption{\label{fig:y-m}\tfa
Mass within an overdensity of 200 versus central y--distortion, for the
219 simulated cluster observations in the dataset.  Each point marks an
observation.  A solid line marks the least--squares best--fit to a power
law,
$y_{0}\,=\,10^{-4.4\pm 0.7}\left(M/10^{15}\msol\right)^{0.98\pm 0.04}$.
A linear relation, expected by theory, is not significantly different;
forcing a fit to a linear relation produces the same coefficient,
and the dashed line on the figure marking this relation cannot be
distinguished from the solid line.}
\end{figure}

While the success of the expected scaling law is quite striking,
the small scatter about the mean relation is also of note.
Modelling the distribution of possible values of the y--parameter
for a cluster of a given mass as log--normal, the simulated
data show a dispersion of $\sigma\,=\,0.15$ in $\log y_{0}$.  While larger than
the scatter predicted by simulations for the $\temp$--$M$
relation (Evrard, Metzler \& Navarro 1996, hereafter \cite{EvMeNa96}), this
dispersion is significantly smaller than the scatter
predicted by the luminosity--mass relation observed in simulations
(\cite{Metz95} 1995), and observed in the luminosity--temperature correlation
for X--ray clusters (\cite{ArEv98} 1998).

The magnitude of the intrinsic scatter about the mean
SZ--mass relation is important.  Since the
predicted abundance of clusters falls with increasing mass,
the net effect of such scatter is to artificially populate the
``bright'' end of the abundance function.  In other words,
such an intrinsic scatter would make the real universe appear
to have more high--mass clusters than are actually present.
Since the high--mass end of the the cluster
abundance function also contains most of the cosmological constraining
power,
a large intrinsic dispersion in the
correlation of an observable with mass makes using the cluster
abundance to constrain cosmological parameters extremely difficult.
Most
efforts to construct the cluster abundance include an attempt
to correct for this.  However, an accurate correction
requires an understanding of the intrinsic dispersion in
the correlation with mass, which has not been available.

Consequently, the comparatively small scatter suggests that
the SZ central decrement is a much more robust observable to
use in constructing an abundance function than X--ray luminosity.
If this result holds at higher redshift as well, then along with
the redshift--independence of the SZ signal, and the expected
self--similar scaling of the SZ signal strength with redshift
at fixed mass, we can expect a large dataset of measurements of
the SZ central temperature decrement to yield a much better
estimate of the high--redshift cluster abundance.

\section{The Cluster $\langle y\rangle _A$--M Relation}

Cluster entropy profiles drawn from simulations such as used here
typically exhibit self--similar behavior outside of the cluster core, but a
dispersion in scaled entropy values in the center of the cluster.
Observations suggest such a dispersion in central entropy
states, as we can demonstrate.
First, note that the cluster luminosity--temperature relation
is both theoretically and empirically well--described by a power
law, $L_{x}\,\propto\,\temp^{\alpha}$.  Second,
the fact that most emission originates
from the cluster core allows us to approximate the X--ray luminosity
by $L_{x}\,\propto\,n_{0}^{2}\ \temp^{1/2}\ r_{c}^{3}$, where $n_{0}$ is
the central gas density and $r_{c}$ is the cluster X--ray core
radius.  Equating, we can solve for the quantity
$s^{\prime}\,=\,\temp/n_{0}^{2/3}$, which is simply related to the
entropy by $s\,=\,\log s^{\prime}$; we obtain
\beq
s^{\prime}\,\propto\,r_{c}\ \temp^{\left(7 - 2\alpha\right)/6}.
\eeq
If the relation between central entropy and cluster mass had little
intrinsic scatter, then the tight mass---temperature relation would
imply a small dispersion in observed values of cluster X--ray core
radius at a given X--ray temperature.  In other words, we should see
a tight relation between X--ray core radius and X--ray temperature;
this is contradicted by observations (\cite{Mohr98} 1998; \cite{Mush98} 1998).
At fixed temperature, thus mass,
there is a wide range of observed core radii, and thus by
inference central entropy states.

Therefore, the result that the magnitude of
the scatter in the cluster y---M relation lies
between that of the $\temp$--$M$ and $L_{x}$--$M$
relations is not surprising.  While the $\temp$--$M$ relation
is expected to be tight (\cite{EvMeNa96}), as the gas specific energy
is determined
by the potential the gas falls through, the dispersion in central
entropy noted above results in a dispersion in central gas density values.
The y--parameter depends upon an integral of the gas density along
the line of sight, so such an intrinsic dispersion introduces scatter
in the correlation with mass.  However, the X--ray emission depends
upon an integral of the square of the density; the intrinsic scatter
in the luminosity---mass relation is correspondingly larger.

That the gas density (and thus gas entropy) appears in simulations to
converge to self--similar behavior at larger radii suggests that the
scatter in correlations between mass and ICM--related properties
could be reduced through the use of a statistic which increases the
weight of data from outside the cluster core.  For X--ray emission,
this has been demonstrated through the prediction and observation
of a tight correlation between cluster temperature and radius
containing a fixed amount of X--ray emission (\cite{MoEv97} 1997).
It should be possible to construct
an SZ observable which correlates well with mass and shows much
smaller intrinsic disperson at fixed mass than the central
y--distortion.

Defining our new statistic as $\langle y\rangle _{A}$, we can write
$\langle y\rangle _{A}$ in terms of the y--distortion induced by a
line--of--sight at angle $\theta$ from the cluster center by
\beq
\langle y\rangle _{A}\,=\,
\int y\left(\theta\right) W\left(\theta\right) \rm{d}^2\theta,
\eeq
where $W\left(\theta\right)$ is a window function that determines
the relative weighting of different parts of the cluster.  The
simplest approach is to use a top--hat window function;
$\langle y\rangle _{A}$ is then simply an average of the SZ map
within some chosen radius.

It is important that the radius used to construct
$\langle y\rangle _{A}$ be a {\em self--similar} radius, \ie\  the radius out
to some fiducial overdensity $\delta$ in 3-D, $r_{\delta}$.
This ensures that the same regions of different clusters are
being examined in constructing
$\langle y\rangle _{A}$.  In practice, however, we would not know the
radius $r_{\delta}$ for an observed cluster.  It could be deduced
from the X--ray temperature; but if we have a good
measurement of the temperature, then we immediately have access to a
mass estimator which is independent of the gas density, and thus
presumably has smaller intrinsic error.  The point is to develop an
SZ--based mass estimator to be used at higher redshifts, where X--ray
temperatures are difficult to obtain.

We solve this problem by {\em estimating} $r_{\delta}$ using the central
value of the y--parameter alone.  In the previous section, we
demonstrated a correlation between the central y--distortion and cluster
mass at $r_{200}$; we use this relation to estimate $r_{200}$,
then use this value to set an outer limit for averaging an SZ map to
construct $\langle y\rangle _{A}$.  Here we will use $0.3\,r_{200}$,
corresponding to $\delta\,\sim\,2000$ for the NFW density profile.
With our previous result of
$y_{0}\,=\,3.98\,\times\,10^{-5}M_{15}^{0.98}$, where $M_{15}$
is the mass within an overdensity of 200 in units of $10^{15}\msol$, and with
$M_{200}\,=\,200\bar{\rho}\,\times\,4\pi r_{200}^3/3$ by definition,
we obtain
$r_{200}^{est}\,\approx\,1.61\ \Omega_{0}^{-1/3}\left(y_{0}/10^{-5}\right)^{1/3} \rm{Mpc}$.
For each mock image, we extract $y_{0}$, use this value to find
$r_{200}^{est}$, and then construct the average
$\langle y\rangle _{A}$ within $0.3\,r_{200}$; this radius would
correspond to
$\theta\,\approx\,1.26^{\prime}$ for $y_{0}\,=\,10^{-5}$ at $z\,=\,0.5$.

The result is shown in Figure 2, with a best--fit power law shown
as a solid line, while the best fit to Eq. (\ref{eq:ymreln}) is
shown as a dashed line.  The two are barely distinguishable.
The slope of the best fit relation is
$0.97\pm0.01$, slightly shallower than the expected slope; the
plot shows, however, that a linear scaling provides a good fit
as well.  More notably, the dispersion in the relation is much
reduced; when modelled as log--normal, the scatter in
$\log \langle y\rangle _{A}$ at fixed mass is only $0.06$
--- that is, $13 - 15$\% in $\langle y\rangle _{A}$ ---
much smaller than for the central decrement alone and comparable
to that found by EMN for the mass--temperature relation at
overdensities of $\ssim 500$.

\begin{figure}[t]
\centerline{\epsfxsize=3.5truein\epsffile{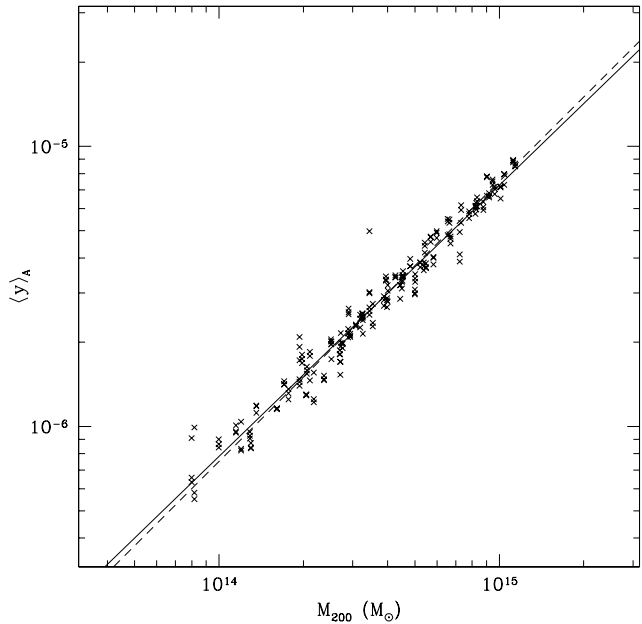}}
\caption{\label{fig:yavg-m}
Mass within an overdensity of 200, $M_{200}$, versus average
y--distortion over a circular area of radius $0.3\,r_{200}^{est}$,
$\langle y\rangle_{A}$.  $r_{200}^{est}$ is an estimate
of the three--dimensional radius containing an overdensity of
200, derived from the central SZ signal $y_{0}$.  Each point marks
one of the 219 simulated cluster observations.
A solid line marks the least--squares best--fit to a power
law,
$\langle y\rangle_{A}\,=\,
10^{-5.1\pm 0.3}\left(M/10^{15}\msol\right)^{0.97\pm 0.01}$.
Forcing a fit to a linear relation produces the same coefficient;
this curve is drawn in a dashed line on the figure.
}
\end{figure}

It may seem puzzling that the scatter in the $\log \langle y\rangle _{A}$ ---
$M$ relation is less than that in the $y_{0}$ --- $M$
relation, given that we use $y_{0}$ to choose a radius within
which to construct $\log \langle y\rangle _{A}$.
However, note first that the dispersion in
$r_{200}^{est}$ will be only $1/3$ the dispersion in
mass inferred from the central signal.  Second, the fact that
the SZ signal should fall with cluster radius suppresses the
error introduced by using an incorrect inferred $r_{200}$.  For
instance, imagine that a cluster with a given mass has
an observed $y_{0}$ higher than expected from the mean
relation; $r_{200}^{est}$ will then be higher than the
correct value.  However, since the signal falls with radius,
constructing an average within this incorrect radius depresses
the result, counteracting the impact of the overlarge central
value.  This works to suppress the dispersion in
$\langle y\rangle _A$ for clusters at a fixed mass that would
be induced by the dispersion in $y_{0}$.

\section{Discussion}

The simulation results discussed above evidence a correlation between
the strength of SZ signal and collapsed mass, in agreement with
theoretical expectation.  The intrinsic scatter observed in the
relation between the central value of the Compton y--parameter $y_{0}$
and mass is greater than seen in the mass---temperature relation,
but considerably smaller than simulations show in the relation between
X--ray luminosity and mass, as expected.  The scatter in this relation can be
reduced to a level near that of the mass --- temperature
relation through the use of statistics which depend
less sensitively on central gas and more sensitively on gas
at larger radii.

Ssome cautionary notes are in order.  First of all, the {\em optimal}
statistic to use will in all likelihood be experiment--dependent,
since different experiments
sample the sky differently.
Also, such a correlation with mass should be affected by limits in
observational resolution; when a cluster is not resolved, the y--parameter
measured will scale with the angular--diameter distance to the
cluster as $y\,\propto\,d_{A}^{-2}$.

There are also issues of concern about the simulations used here.
Chief among these are numerical resolution limitations, which likely
serve to reduce the dispersion in the SZ --- mass correlation.
Furthermore, the effect of missing physics which breaks self--similarity,
such as cooling or the inclusion of supernova--driven galactic winds,
must be considered.  Next, the simulated clusters used in this
investigation were taken at $z\,=\,0$; numerical resolution issues
are of greater
concern as redshift increases.  It would be worthwhile to investigate
in detail, using higher--resolution simulations, the status of the
SZ --- mass correlation at redshifts of $0.5$ or 1.
Finally, the effects of the choice of background cosmology ---
particularly, the values chosen for $\Omega_{0}$ and the
baryon fraction $f_b\,=\,\Omega_b/\Omega_{0}$ --- must be examined.

One hope for verifying or calibrating the SZ --- mass
correlation comes from comparing SZ--derived masses with those from
X--ray observations or from
gravitational lensing.  Since low redshift clusters typically
have better temperature estimates, the planned
Viper Sunyaev--Zel'dovich Survey (\cite{Rome98} 1998)
should provide a good opportunity to test this correlation.

Future work is planned to study these issues, as well as to examine
the systematics of using SZ observations to constrain the cluster baryon
fraction.

\bigskip
This research was supported at the University of Illinois by the
NSF, and with the NASA/Fermilab Astrophysics Group at the Fermi
National Accelerator Laboratory under NASA grant NAG 5--7092.  The
author would like to thank John Carlstrom, Bob Nichol, and Martin
White for useful discussions, and the Aspen Center for Physics,
where a large portion of the work was done.

\clearpage

\end{document}